\documentclass[a4paper, 10pt]{article}
\usepackage[margin=2.5cm]{geometry}

\usepackage{adjustbox}
\usepackage{graphicx, cite, booktabs, url, tabularx}
\PassOptionsToPackage{hyphens}{url}\usepackage{hyperref}
\hypersetup{colorlinks=true, citecolor=blue, linkcolor=black, urlcolor=blue}
\usepackage[labelfont=bf,labelsep=period,justification=justified, singlelinecheck=false]{caption}
\usepackage{amssymb, amsmath, bm}
\graphicspath{ {./figures/} }
\usepackage{seqsplit}
\newcommand{\ttb}[1]{\texttt{\seqsplit{#1}}}
\newcommand{\hchan}[1]{%
  {\setlength{\fboxsep}{1pt}
   \setlength{\fboxrule}{0.4pt}
   \fbox{\textbf{#1}}%
  }%
}

\title{\textbf{An Open-Access Multi-modal Dataset for Cognitive, Motor and Cognitive-Motor Tasks}}

\author{Zaineb Ajra\textsuperscript{1*}, Grégoire Vergotte\textsuperscript{1*}, Stéphane Perrey\textsuperscript{1}, Lilian Evra\textsuperscript{1},\\ Simon Pla\textsuperscript{1}, Gérard Dray\textsuperscript{1}, Jacky Montmain\textsuperscript{2}, Binbin Xu\textsuperscript{1}
}
\date{
\small
\textsuperscript{1}{EuroMov Digital Health in Motion, Univ Montpellier, IMT Mines Ales, France}\\
\textsuperscript{2}{SyCoIA, IMT Mines Ales, Ales, France}\\
corresponding: \texttt{binbin.xu@mines-ales.fr}\\
\textsuperscript{*}{Zaineb Ajra and Grégoire Vergotte contributed equally to this work.}
}

\begin{document}

\maketitle

\begin{abstract} 
Open neuroimaging datasets are important for developing reusable analysis methods, but many available resources focus on isolated tasks or single modalities. Here, we describe a multimodal dataset acquired from 30 healthy adults during three sessions of seven cognitive, motor and combined cognitive-motor tasks. The protocol includes N-back, mental arithmetic, passive arm movement, motor imagery, active motor execution, N-back arithmetic and a full cognitive-motor condition. The dataset contains synchronized electroencephalography, functional near-infrared spectroscopy, electrocardiography, behavioral responses and subjective ratings. Data are organized in Brain Imaging Data Structure format and are publicly available for reuse. The hierarchical task design allows users to compare isolated and combined task conditions, while repeated sessions support analyses of intra-participant variability. Technical validation focuses on EEG and fNIRS responses and on subjective task ratings. The dataset may support research in multimodal neurophysiology, cognitive-motor interaction, neuroergonomics and brain-computer interface methods.
\end{abstract}

\section{Background \& Summary} 

Neuroscience has made significant progress in recent decades, largely due to the development of new neuroimaging techniques and the availability of openly shared datasets. These advances complement behavioral studies and provide new insights into the brain-behavior relationship, but also introduce challenges, particularly in designing more ecologically valid experimental paradigms and developing analytical methods capable of handling the complexity of multimodal data. Recent reviews of mobile and ecological neuroimaging show that moving from highly controlled laboratory tasks to more naturalistic behaviours makes the experiment more complex, for example for sensor integration, movement artefacts, synchronization and signal interpretation \cite{Stangl2023Mobile,Richer2024Mobile}.

Electroencephalography (EEG) and functional near-infrared spectroscopy (fNIRS) are two powerful neuroimaging methods used to study brain activity both in controlled laboratory settings and in real-life contexts. These methods have been applied to investigate neurophysiological adaptations under task constraints in real-world activities \cite{seidel2021task,carius2023increased,tamburro2023ecological,jacobsen2022mobile,mavros2022mobile,slutter2021exploring,cao2019multi,fresnel2021cerebral}. These studies demonstrate the potential of EEG and fNIRS in real-world scenarios. However, challenges remain in integrating and standardizing these methods for widespread use across diverse naturalistic environments.

Several open-access datasets using EEG or fNIRS have been proposed to study cognitive or motor tasks. For example, such datasets have been used to examine mental states \cite{hinss2023open}, resting state and cognitive states \cite{wang2022test}, motor imagery \cite{ma2022large,dreyer2023large} and motor tasks \cite{cockx2023dealing}. Other public datasets have focused on isolated paradigms, such as N-back, discrimination/selection-response or word-generation tasks \cite{Shin2018Simultaneous}, Stroop tasks \cite{Chen2023Open}, unilateral finger and foot tapping \cite{Bak2019Open}, or passive auditory stimulation \cite{Luke2021Analysis}. Here, isolated task means a condition mainly related to one functional domain, for example a cognitive task without overt motor component, or a motor task without concurrent cognitive demand. These datasets are useful to study specific processes, such as memory, attention, vigilance, task-switching, motor imagery or limb movement. However, as they usually focus on one main task family or one acquisition context, they are less adapted to examine cognitive, motor and combined cognitive-motor demands in the same repeated-session protocol.

Although EEG and fNIRS are used to study distinct neural processes, with different temporal and spatial resolution properties, their combination offers new insights into the neural mechanisms underlying cognitive-motor tasks \cite{su2023simultaneous}. Studies integrating EEG and fNIRS have also opened the way to the development of new multivariate analytical methods, allowing the combination of these complex neurophysiological signals for improved interpretability \cite{li2022concurrent}. Nevertheless, there is a lack of publicly available datasets that combine both modalities; most studies rely on either modality in isolation \cite{Chen2023Open}. EEG-based brain-computer interface studies also imply the importance of reliable brain-signal acquisition and processing for human-computer interaction and assistive applications \cite{Sibilano2023BrainComputer}. At the same time, simultaneous EEG-fNIRS acquisition is technically demanding. This requires compatible hardware, trained personnel, careful sensor placement, reliable synchronization and adapted signal processing for modality-specific artefacts \cite{Uludag2014General}. The present dataset should therefore be considered as a reusable multimodal resource that users can inspect and preprocess according to their own methodological objectives.

Electrocardiography (ECG) provides a complementary measure of cardiac activity alongside EEG and fNIRS. Synchronized ECG enables the extraction of heart rate and heart-rate variability, which can provide physiological context for cognitive workload or stress, motor-related autonomic responses, and combined cognitive-motor task demands \cite{arquilla2022utility,Dybvik2025fNIRS}. These cardiac measures complement, rather than directly measure, neural activity, and their sensitivity may vary across tasks. In the present dataset, ECG is therefore included as an auxiliary physiological stream for future reuse, whereas the technical validation focuses on EEG and fNIRS.

Combined cognitive-motor tasks refer here to conditions where cognitive and motor demands are present in the same task block. Multitask datasets are needed to explore relationships between different brain-behavior functional organizations involved in cognitive-motor tasks and to understand how these processes interact in real-world scenarios. Creating a continuum from lab-based single-task designs to multitask and more naturalistic experiments enables EEG and fNIRS methods to be tested step by step while preserving interpretability (i.e. neurophysiologically hypothesis driven) and challenging their robustness and transferability in real-world environments. 

In clinical, rehabilitation and training contexts, cognitive and motor abilities are often evaluated with standardized scales or observation-based instruments. These tools are essential in practice, but they can give rather categorical summaries of performance. Neurophysiological recordings such as EEG and fNIRS may complement these evaluations by giving continuous signals related to cognitive and motor engagement, workload and skill acquisition \cite{Ronca2026novel}. The present dataset is not a diagnostic, clinical validation or training-intervention cohort. It provides a healthy-participant resource for developing and testing multimodal analysis pipelines under controlled cognitive, motor and cognitive-motor task conditions.

We therefore propose a new open access dataset built around a hierarchical ramification of seven tasks (Figure~\ref{fig:exp_paradigm}). The task tree starts from two cognitive tasks (N-back and mental arithmetic) and two motor tasks (passive arm movement and motor imagery) and extends to combined conditions that integrate the following roots: N-back arithmetic, active motor and N-back arithmetic combined with a motor task. This hierarchical structure makes it possible for users to test multiple hypotheses about cognitive-cognitive, motor and cognitive-motor interactions and to examine how isolated and combined processes are represented in the brain across increasing levels of task difficulty. The main contribution of this dataset is the combination of four elements in the same resource: multimodal acquisition, including EEG, fNIRS, ECG, behavioral responses and subjective ratings; a hierarchical protocol with isolated cognitive tasks, isolated motor tasks and combined cognitive-motor tasks; recordings repeated across three sessions; and public release using the Brain Imaging Data Structure (BIDS). 

The resulting raw multimodal psycho-neurophysiological dataset, comprising EEG, fNIRS, ECG, questionnaires and behavioral measures, is publicly available in BIDS format \cite{gorgolewski2016brain}. The technical validation reported here focuses on EEG, fNIRS and subjective ratings; the scope of the ECG stream is further specified in the Usage Notes.

\subsection{Related public datasets and scope of the present dataset}
We compared the present resource with public datasets sharing at least one aspect with our work: EEG or fNIRS acquisition, multimodal neurophysiological recordings, cognitive tasks, motor tasks, repeated recordings or public reuse (Table~\ref{tab:related_datasets}).

This comparison is made to position the dataset. It is not a benchmarking analysis against equivalent datasets. Such equivalent datasets are difficult to identify because multimodal resources differ in modalities, task structure, ecological level, population, number of sessions and data organization.

\begin{table}[!htb]
\small
\centering
\caption{Comparison with related public neurophysiological datasets. The table focuses on dataset characteristics useful for reuse, not on benchmark performance.}
\label{tab:related_datasets}
\begin{adjustbox}{max width=\linewidth}
\begin{tabular}{p{2.2cm}p{2.2cm}p{2.2cm}p{2.5cm}p{1.6cm}p{5.5cm}}
\toprule
Dataset & Modalities & Participants & Task focus & Cognitive-Motor task & Main difference with the present dataset \\
\midrule
Shin et al. \cite{Shin2018Simultaneous} & EEG, NIRS & 26 healthy adults & N-back, DSR, word generation & No & Multimodal cognitive-task dataset, but without motor tasks or combined cognitive-motor task hierarchy. \\
Chen et al. \cite{Chen2023Open} & EEG, fNIRS & 21 healthy adults & Stroop task & No & Dual-modality Stroop dataset, but focused on one cognitive paradigm. \\
Bak et al. \cite{Bak2019Open} & fNIRS & 30 healthy adults & Finger and foot tapping & No & Motor fNIRS dataset, but without EEG, ECG, repeated sessions or cognitive-motor combinations. \\
Luke et al. \cite{Luke2021Analysis} & fNIRS & 17 human listeners & Passive auditory stimulation & No & Useful fNIRS resource for auditory responses, but not designed for motor or cognitive-motor interactions. \\
Present dataset & EEG, fNIRS, ECG, behavior, subjective ratings & 30 healthy adults & Seven cognitive, motor and combined tasks & Yes & Hierarchical protocol with isolated and combined cognitive-motor tasks, synchronized multimodal recordings and BIDS organization. \\
\bottomrule
\end{tabular}%
\end{adjustbox}
\end{table}

The remainder of the article is organized as follows. The Methods section describes participants, experimental paradigm, acquisition systems and synchronization. The Data Records section describes the released files and metadata. The Technical Validation section reports validation analyses. Finally, the Usage Notes section gives practical information for reuse. 

\section{Methods}
\label{sec:methods}

\subsection{Ethics statement}
The study was performed in agreement with the standards set by the Declaration of Helsinki (2013) involving human participants. The protocol was reviewed and approved by the local research Ethics Committee (IRB-EM 1912B, EuroMov, Montpellier, France). Thirty healthy right-handed adults (12 males, 18 females, 25.5 $\pm$ 5.4 years), without known neurological disorders, were recruited by word of mouth within the laboratory and on the University campus. Before inclusion, participants received oral and written information about the study objectives, experimental procedures, recorded data streams, expected duration, potential discomforts, confidentiality measures, their right to withdraw, and the intended scientific use and sharing of anonymized data. Written informed consent was obtained from each participant before any study procedure. The consent form specified that anonymized data could be used for scientific publications and shared with the scientific community. Participants are identified only by BIDS-compliant codes ranging from \ttb{sub-001} to \ttb{sub-030}. All raw data were converted to the Brain Imaging Data Structure by the same experimenter using these participant codes. In accordance with the data-protection plan approved by the ethics committee, identifiable consent records and other participant-identifying information were stored separately from the research data in a locked room, and access was restricted to the Principal Investigator.

\subsection{Experimental paradigm}
\begin{figure}[htb]
\centering
\includegraphics[width=\textwidth]{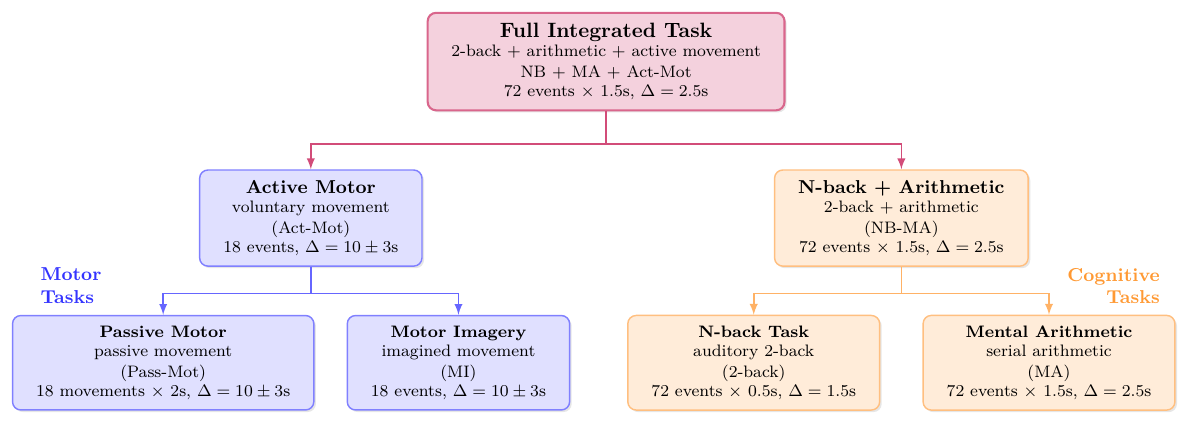} 
\caption{Experimental paradigm overview}
\label{fig:exp_paradigm}
\end{figure}

The experimental paradigm is illustrated in Figure~\ref{fig:exp_paradigm}. Each participant completed one 90-minute visit in a quiet, isolated room. Before the experiment begun, we ensured optimal conditions (e.g., no outside noise, phones switched off). The experiment consisted of three sessions during which participants performed seven different tasks designed to evaluate both cognitive and motor functions: Mental Arithmetic (MA), N-back (NB), Motor Imagery (MI), Passive Motor (Pass-Mot), Active Motor (Act-Mot), N-back Arithmetic (NB-MA) and Full Task (NB-MA-Act-Mot). These tasks were presented in a randomized order, with each session lasting approximately 30 minutes, including both task and post-task questionnaires. 

Before data acquisition began, participants were briefed on the study procedures and completed a short questionnaire based on the Edinburgh Handedness Inventory \cite{veale2014edinburgh} to confirm right-handedness. Participants were then equipped with EEG-fNIRS neuroimaging systems while receiving task instructions. 

In addition to neurophysiological data, we collected ratings of sleepiness level before each session using the Karolinska Sleepiness Scale (KSS) \cite{kaida2006validation} and also ratings of cognitive load after each task using a nine-point Likert scale \cite{ouwehand2021measuring}.

The recording room was dimly lit and each participant sat comfortably in a chair on a Biodex isokinetic dynamometer in a standardized position, with eyes directed towards a screen displaying the instructions. {Short inter-session breaks, generally around 5 minutes or more, were allowed to reduce fatigue and limit carry-over effects across repeated cognitive, motor and combined task blocks.}

\paragraph{Experimental conditions:}
the experimental paradigm consisted of seven consecutive tasks, performed in a randomised order and each lasting three minutes. There was a 30-second rest period between each task. This sequence was repeated three times, with at least 2 minutes of rest between repetitions. These rest periods were included to support participant compliance and limit immediate carry-over between task conditions, as considered in repeated motor-task protocols \cite{Rinaldi2026Towards}.

\begin{enumerate}
    \item \textbf{Mental arithmetic task (MA):} Participants heard digits between 0 and 9 and had to perform simple additions or subtractions such that the outcome remain within the 0-9 range. Each arithmetic operation was considered an event and lasted 1.5 seconds. In total, 72 events were presented, with a new event occurring every 2.5 seconds. This task constitutes a classic cognitive load manipulation targeting working memory processes \cite{destefano2004role}.

    \item \textbf{N-back task (NB):} A 2-back version of the N-back task was used to impose a moderate level of cognitive workload and avoid cognitive overload. Numbers between 0 and 9 were presented auditorily and participants were instructed to press a button when a 2-back target occurred (18 targets in total). Each number was considered an event and lasted 0.5 seconds. In total, 72 events were presented, with a new event occurring every 1.5 seconds. 

    \item \textbf{Motor imagery task (MI):} Participants were asked to imagine performing a movement with their right arm without actually moving or contracting their muscles when they heard a beep. Each auditory stimulus was presented at intervals of $10 \pm 3$ seconds, with 18 targets throughout the task.

    \item \textbf{Passive motor task (Pass-Mot):} Participants positioned their right arm on a Biodex robotised dynanometer, which performed the movement for them. The arm was abducted, with the forearm flexed to 90 degrees in the transverse plan. The robotic arm rotated  internally and externally over a range of 60 degrees of motion. Each movement lasted 2 seconds and was spaced at random intervals of $10 \pm 3$ seconds, 18 targets throughout the task.

    \item \textbf{Active motor task (Act-Mot):} For each beep sound, participants were instructed to perform the task by voluntarily moving (range 60 degrees at a moderate speed) the Biodex robotised dynanometer arm. Auditory stimuli were presented at intervals of $10 \pm 3$ seconds, with 18 targets in total throughout the task.

    \item \textbf{N-back arithmetic task (NB-MA):} Participants were required to press a push button when the sum of the numbers, presented auditory and ranging from 0 to 9, corresponded to the 2-back condition. This task aimed to test participants' ability to process two types of mental information simultaneously. Each event lasted 1.5 seconds, with a total of 72 events presented, with a new event occurring every 2.5 seconds (18 targets).

    \item \textbf{Full task (NB-MA-Act-Mot):} Participants were required to move their arm (as for 5. Act-M) when the NB-MA condition occurred, instead of pressing a push button. In other words, they had to move their arm when the sum of the numbers, presented auditorily and ranging from 0 to 9, corresponded to the 2-back condition. This task integrates cognitive (NB and MA) with an active motor (Act-Mot) aspects. It represents the most difficult condition simulating real-world scenarios where cognition and movement must be coordinated simultaneously. Each event lasted 1.5 seconds, 72 events were presented, with a new event occurring every 2.5 seconds (18 targets).
    
\end{enumerate}

Stimuli were delivered using Psychtoolbox (V3.0.19), a MATLAB-based toolbox; auditory beeps (generated as a sine wave in MATLAB and presented for 500 ms) and synthesized speech (generated using Espeak (v1.48.04) combined with Mbrola French voice (fr1)) were used.

\subsection{Data acquisition}

\begin{figure}[h]
\centering
\includegraphics[height=5cm]{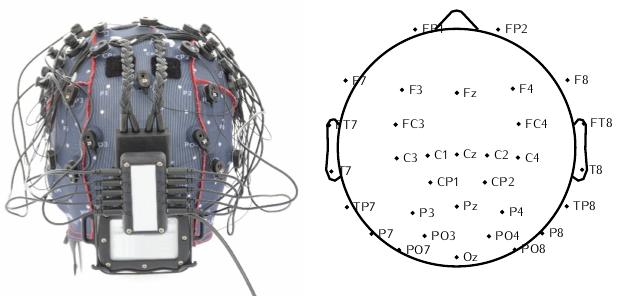}
\caption{Configuration of the experimental fNIRS-EEG system. Left: picture of the head cap (rear view). Right: spatial layout of the 32-channel EEG electrodes arranged according to the international 10-20 system.}
\label{fig:x EEG position}
\end{figure}

\paragraph{Neurophysiological data acquisition}
Participants wore a headset equipped with an integrated EEG-fNIRS system to simultaneously monitor electrical brain activity (EEG) and hemodynamic responses (fNIRS) throughout all tasks.
The device configuration is shown in Figure \ref{fig:x EEG position}.

\begin{itemize}
    \item \textbf{EEG system:} EEG data were acquired using a 32-channel gel electrode cap (standard 10/20 system) with a g.tec biopotential amplifier (g.Nautilus PRO Headset, g.tec medical engineering GmbH, Austria). Data were sampled at 250 Hz. An ear-clip electrode, attached with gel, was used as the reference. The device includes an internal impedance check and electrode-skin-impedance was kept below 5 k$\Omega$ before starting the experiment. Data were recorded continuously during each session and no filtering was applied during the acquisition. Participants were instructed to minimize movement during the experiment. 

    \item \textbf{fNIRS system:} The fNIRS montage includes two Octamon sytems resulting in 16 channels covering Prefrontal cortex and Sensorimotor cortex.
    \begin{enumerate}
        \item Prefrontal cortex (PFC): fNIRS data were recorded using an Octamon system (Artinis Medical Systems, the Netherlands) composed of 8 sources (760 and 850 nm) and 2 detectors, forming 4 channels per hemisphere (inter-optode distance 35 mm). Raw intensity data were recorded at 10 Hz using Oxysoft (v3.2.72). 

        \item Sensorimotor cortex (SMC): fNIRS data were collected using a Octamon+ system (Artinis Medical Systems, the Netherlands) composed of 8 transmitters with two wavelengths (760 and 850 nm). The 8 transmitters and 2 receivers (inter-optode distance 30 mm) resulted in $2 \times 4$ fNIRS channels covering left and right SMC. Raw intensity data were recorded at 10 Hz simultaneously with EEG using Simulink models provided by g.tec.
    \end{enumerate}

\end{itemize}

\paragraph{Physiological data acquisition}
Physiological data (ECG) were recorded using the Delsys Trigno wireless system to capture cardiac activity during the tasks. 

Although the physiological recordings were acquired with skin-mounted Delsys sensors fixed before the tasks, contact-related artefacts cannot be fully excluded in such recordings, as perspiration or small sensor-skin displacements may affect signal quality \cite{Abdoli-Eramaki2012effect,Suglia2026Muscle}. The present validation therefore does not assume that the physiological channels are artefact-free, and modality-specific quality control should be applied before analyses of these signals.

\begin{itemize}

     \item \textbf{Electrocardiography data (ECG):} ECG data were recorded using an EKG biofeedback sensor from the Trigno wireless EMG system (Delsys, USA) at 1\,986 Hz. Sensors were placed below the left and right pectoralis major muscles. The ECG was used to monitor the cardiac activity of participants to detect potential psychological stress responses during the experimental conditions or to regress to other signals \cite{arquilla2022utility}. The ECG stream is included as additional physiological signal. Users interested in cardiac measures or physiological-confound analyses should perform dedicated ECG quality control before extracting heart-rate or heart-rate-variability measures.

\end{itemize}

\paragraph{Behavioral data acquisition}
\begin{itemize}
    \item \textbf{Push button:} During the NB and NB-MA tasks, participants pressed a push-button when a 2-back target was detected or when the sum of the numbers corresponded to the 2-back condition. The push-button signal was sampled at 148.13 Hz.

    \item \textbf{Isokinetic device:} A Biodex System 3 (Biodex Medical Systems, Inc., United States) was used to measure torque during passive movements (Torque = 10 N$\cdot$m) and voluntary movements (isokinetic mode: 0.5 N$\cdot$m). This system allows to record force output at a constant movement velocity during the tasks.
\end{itemize}

\paragraph{Subjective data acquisition}
\begin{itemize}
    \item \textbf{Demographic data:} Demographic questionnaires were coded in MATLAB R2021b and included questions on highest education degree, date of birth, gender, handedness and date of the experiment. 

    \item \textbf{Edinburgh handedness inventory:} Participants completed a simplified form of the Edinburgh Handedness Inventory to confirm their dominant hand. The questionnaire includes four activities (writing, throwing an object, brushing teeth and using a spoon), each rated on a 5-point scale: ``Always on the left, Usually on the left, both equally, Usually on the right, Always on the right''.
    
    \item \textbf{Karolinska Sleepiness Scale (KSS):} The KSS is a 9-point self-report scale used to access the current level of sleepiness, ranging from ``extremely alert'' to  ``extremely sleepy, fighting sleep'' \cite{kaida2006validation}. 

    \item \textbf{Likert Rating Scale:} Perceived task difficulty was assessed using a 9-point Likert scale frequently used in Cognitive Load Theory research, along with a Visual Analogue Scale ranging from 0 to 100\%. Participants rated their perceived mental effort after each task, from ``very very low mental effort'' to ``very very high mental effort'' \cite{ouwehand2021measuring}.
\end{itemize}

\subsection{Data synchronization}

\begin{figure}[!htb]
\centering
\includegraphics[width=\textwidth]{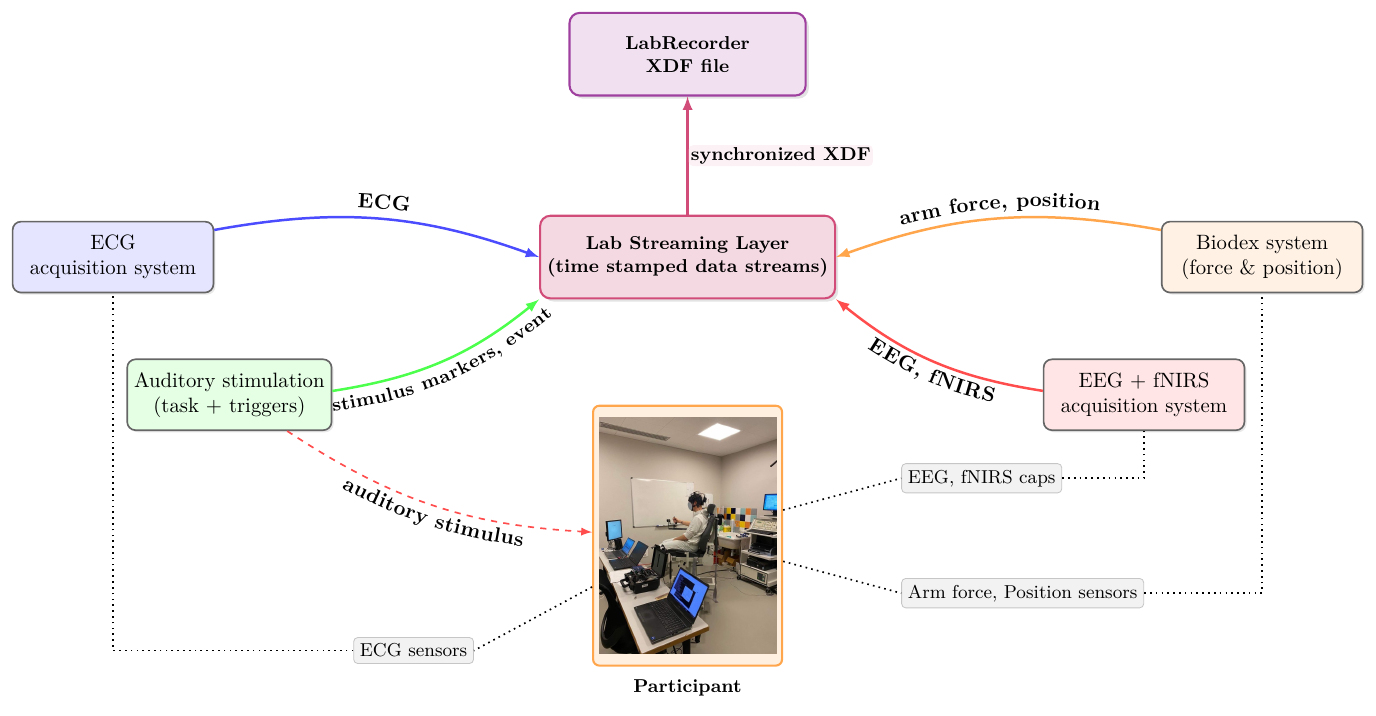}
\caption{{Overview of the synchronized data acquisition setup using Lab Streaming Layer. Arrows indicate individual streams.}}
\label{fig:data_sync_setup}
\end{figure}

During each of the three sessions, we collected multiple data streams from each participant, including neurophysiological data (EEG, fNIRS), physiological data (ECG), behavioral data (reaction times from the push button; arm force and position measures from the Biodex dynanometer) and subjective data (questionnaires). To ensure precise synchronization and alignment across the experimental timeline, we used LabRecorder from the Lab Streaming Layer (LSL) framework (\url{https://github.com/sccn/labstreaminglayer}) to record both data streams and stimulus markers. The LSL output was saved as an XDF file containing all synchronized streams, ensuring that neurophysiological, physiological and behavioral data were properly aligned. {We kept the native sampling rate of each acquisition stream rather than resampling all modalities before sharing the dataset. This avoids fixing one preprocessing choice for all future analyses, especially for ECG or other physiological signals. One can therefore resample the signals according to their own analysis pipeline.} The complete data acquisition / synchronization setup is illustrated in Figure \ref{fig:data_sync_setup}.

\section{Data Records}
\label{sec:data_records}

\begin{figure}
    \centering
    \includegraphics[width=0.95\linewidth]{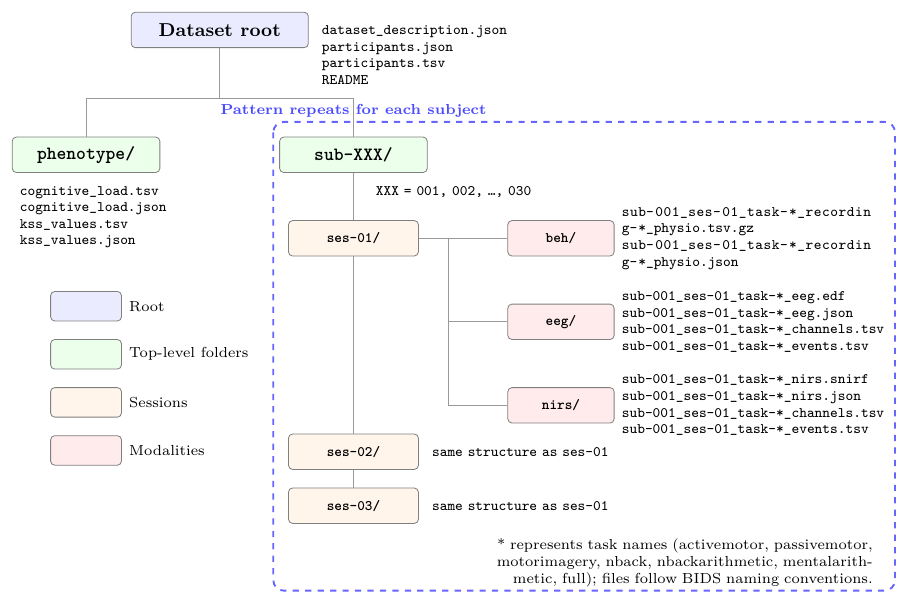}
    \caption{{Overview of the dataset structure}}
    \label{fig:datafolder_structure}
\end{figure}

An overview of the repository \cite{ds007554} organisation is shown in Figure~\ref{fig:datafolder_structure}. At the dataset root, we provide the standard BIDS metadata files (\ttb{dataset\_description.json}, \ttb{participants.tsv/json}). The questionnaires data (cogntive load and KSS scores) is placed in the folder \ttb{prenotypes}. For each participant (\ttb{sub-XXX}) and each of the three sessions (\ttb{ses-01}, \ttb{ses-02}, \ttb{ses-03}), data are stored in modality-specific folders (\ttb{beh}, \ttb{eeg}, \ttb{nirs}). Within each session directory, files follow BIDS naming conventions that encode subject, session, task and recording type. For example, EEG and fNIRS files follow patterns such as
\ttb{sub-001\_ses-01\_task-nback\_eeg.edf} and
\ttb{sub-001\_ses-01\_task-nback\_nirs.snirf}, respectively, with associated sidecar files sharing the same stem. Physiological and behavioural recordings follow analogous patterns, for instance
\ttb{sub-001\_ses-01\_task-nback\_recording-ECG\_physio.tsv.gz}.

Tasks are indicated by the \ttb{task-} label (e.g. \ttb{activemotor}, \ttb{passivemotor}, \ttb{motorimagery},
\ttb{nback}, \ttb{nbackarithmetic}, \ttb{mentalarithmetic}, \ttb{full}) and recording modalities are indicated
by the \ttb{recording-} label for physiological files (e.g. \ttb{ECG}, \ttb{biodex}, \ttb{pushbutton}).
Within each modality-specific subfolder, all files are named using BIDS-compliant patterns that encode the subject,
session, task, and, when relevant, the recording type.

Data availability is not identical for all modalities and validation analyses. Supplementary Table~\ref{tab:data_completeness_summary} gives an overview of participant- and modality-level availability, so that users can select the subset adapted to modality-specific or multimodal analyses. This table describes available recordings and validation inclusion, but it does not define the statistical sample size for a specific reuse analysis.

The \texttt{events.tsv} files describe task timing with the columns \texttt{onset}, \texttt{duration}, and \texttt{trial\_type}. In the representative files inspected here, \texttt{trial\_type} uses the labels \texttt{start}, \texttt{target}, and \texttt{trigger}. The \texttt{target} and \texttt{trigger} labels distinguish target and non-target stimulus events in the N-back, N-back arithmetic, and full task files. Event counts can differ between files according to task structure and available recordings. Correctness, response timing and other behavioral variables should be obtained from the behavioral files when available, and not from the event labels only.

\subsection{EEG data}

Raw EEG recordings are stored in the \ttb{eeg} folder as \ttb{.edf} files, with one file per task and session (for example \ttb{sub-001\_ses-01\_task-nback\_eeg.edf}). For each EEG recording, we provide the following BIDS sidecar files sharing the same filename stem:
\begin{itemize}
    \item \ttb{\_eeg.json}, a JSON file describing acquisition parameters (for example amplifier type, sampling frequency, filter settings and reference),
    \item \ttb{\_channels.tsv}, a tabular file listing channel names, types and other channel-specific properties,
    \item \ttb{\_events.tsv}, a tabular file specifying task events, including onset times, durations and event labels.
\end{itemize}
No online filtering was applied during acquisition beyond the settings required by the recording hardware and no offline preprocessing has been applied to the shared EEG recordings. The distributed EEG data therefore correspond to the raw continuous signals exported from the acquisition system, together with the original event markers.
\subsection{fNIRS data}

fNIRS recordings and events are stored in the \ttb{nirs} folder in \ttb{.snirf} format, again with one file per task and session (for example \ttb{sub-001\_ses-01\_task-nback\_nirs.snirf}). For each fNIRS file, we provide:
\begin{itemize}
    \item \ttb{\_nirs.json}, a JSON sidecar that documents acquisition parameters,
    \item \ttb{\_channels.tsv}, a tabular file describing optode and channel properties, including channel indices, source-detector pairs, nominal wavelength and unit, 
    \item \ttb{\_events.tsv}, a tabular file listing experimental events with their onset and condition label.
\end{itemize}

The fNIRS montage includes prefrontal and sensorimotor channels as detailed in the Methods section. The shared \ttb{.snirf} files contain the raw intensity signals as exported from the acquisition software, without additional offline preprocessing.

\subsection{Physiological and behavioural data}

Physiological and behavioural signals are stored in the \ttb{beh} folder as compressed tab-separated value files (\ttb{.tsv.gz}) with associated JSON sidecar files that describe the recorded signals and their units. Filenames encode the subject, session, task and recording type using the \ttb{\_recording-<label>\_physio} key-value pair.

For example, for \ttb{sub-001} in \ttb{ses-01}:
\begin{itemize}
    \item ECG recordings for each task are stored as \ttb{sub-001\_ses-01\_task-<task>\_recording-ECG\_physio.tsv.gz} with accompanying \ttb{sub-001\_ses-01\_task-<task>\_recording-ECG\_physio.json} files that specify sampling frequency, channel names and units.
    
    \item Force and torque measurements from the dynamometer are stored as \ttb{\_recording-biodex\_physio.tsv.gz} with matching JSON sidecars.
    
    \item EMG recordings for the passive motor task are stored as \ttb{sub-001\_ses-01\_task-passivemotor\_recording-EMG\_physio.tsv.gz} with an associated \ttb{\_recording-EMG\_physio.json} file.
    
    \item Push-button signals for the n-back and n-back arithmetic tasks are stored as \ttb{\_recording-pushbutton\_physio.tsv.gz} files with corresponding JSON files that document the mapping between button channels and response types.
\end{itemize}

Each \ttb{.tsv.gz} file contains one column per recorded channel and, when applicable, additional columns that encode task-related variables (for example trial codes or condition markers). The associated JSON files provide the required BIDS metadata so that physiological and behavioural recordings can be interpreted and processed in conjunction with the EEG and fNIRS data.

\subsection{Data completeness and sampling differences}

Data from participant \ttb{sub-008} were excluded due to incomplete sessions. For participants \ttb{sub-010} to \ttb{sub-014}, EEG signals were not usable because of technical issues with the EEG device and their EEG data are therefore not provided. All other modalities for these participants are present in the repository. Participants \ttb{sub-017} and \ttb{sub-019} are kept in the repository, but they were not included in the EEG technical validation after quality-control checks. Users should distinguish between data availability and inclusion in the validation analyses.

For fNIRS signals, data from participants \ttb{sub-001} to \ttb{sub-020} were recorded at 10 Hz, whereas data from participants \ttb{sub-021} to \ttb{sub-030} were recorded at 50 Hz. For the technical validation analyses, the 50 Hz recordings were resampled to 10 Hz to use the same preprocessing settings across participants.

\section{Technical Validation}
\label{sec:technical_validation}

In the technical validation analyses, we focused on EEG and fNIRS signals, together with the subjective ratings collected during the protocol. The aim was to show that these recordings can be used to characterize task-related responses, and that the subjective ratings follow the expected task structure. The validation was not intended to define one mandatory preprocessing pipeline. We therefore report the preprocessing choices used here and representative EEG, fNIRS and subjective-rating analyses. ECG data are included in the dataset for future physiological analyses, but they were not used in the present technical validation.

\subsection{EEG and fNIRS preprocessing}

\paragraph{EEG preprocessing} 
EEG pre-processing was conducted using EEGLAB toolbox \cite{delorme2004eeglab}. The pipeline followed a two-step approach. First, a data sanity check was conducted using the PREP pipeline \cite{bigdely2015prep} to ensure data quality for further analysis. Each dataset was inspected individually. Participants \ttb{sub-017} and \ttb{sub-019} were excluded from the EEG technical validation after this step, but not from the complete repository.

For the remaining participants, raw EEG data from the first session were high-pass filtered at 0.1 Hz, followed by line-noise removal using the \ttb{cleanline} function (power line frequencies 50 and 100 Hz, sliding window step 1 s). Artefact subspace reconstruction was then applied using \ttb{clean\_rawdata} function with default parameters to remove flat-line channels and correct artefacts. Bad channels were interpolated using spherical interpolation and data were re-referenced to average reference. 

Finally, a low-pass filter at 30 Hz was applied to reduce high-frequency noise and muscle artefacts. No ECG-specific artefact correction was applied in the EEG technical-validation pipeline. The simultaneously recorded ECG stream was not used as a reference signal for EEG cleaning in the present analyses. Therefore, the EEG validation results should be interpreted as representative of the stated preprocessing pipeline rather than as outputs of a cardiac-artifact-optimized pipeline. Users performing analyses that are sensitive to cardiac contamination should inspect possible ECG-related EEG components and may use the synchronized ECG recordings to apply dedicated correction approaches, such as ICA-, template- or regression-based methods. The resulting preprocessed data were segmented into epochs using time intervals from -0.5 to 2 s relative to event onset and baseline correction was applied using the interval from -0.5 to 0 s.

\paragraph{fNIRS preprocessing}
Raw fNIRS data were preprocessed using the Homer3 MATLAB toolbox \cite{huppert2009homer}. First, signal quality was assessed using the Scalp Coupling Index (SCI; threshold defined as the median of windows with SCI $> 0.5$) implemented in the qt-nirs toolbox (\url{https://github.com/lpollonini/qt-nirs}) with the following parameters: band-pass filter 0.5 to 2.5 Hz, window length 5 s and no overlap. Channels that did not meet this criterion were excluded from subsequent analyses. 

The remaining raw intensity data were converted to optical density and motion artefacts were corrected using wavelet-based correction (interquartile range 1.5). A band-pass filter between 0.009 and 0.7 Hz was then applied before converting optical density to relative concentration changes using the modified Beer-Lambert law (partial pathlength factor [1, 1]). Concentration changes in oxyhaemoglobin (O2Hb) and deoxyhaemoglobin (HHb) were epoched from -2 to 5 s relative to event onset and baseline corrected using the interval from -2 to 0 s.

\subsection{Subjective data analysis}

To assess whether participants perceived task difficulty consistently across the seven experimental conditions, we analyzed the results of the likert scale ratings using a linear mixed-effects model fitted by restricted maximum likelihood. Task and session were included as fixed effect and Subjects as a random intercept, which allowed us to account for missing ratings (e.g., one participant for the passive motor task).

\begin{table}[htb]
\centering
\small
\caption{Pairwise post hoc comparisons of perceived task difficulty between all task pairs. The linear mixed-effects model included \textit{Task} as a fixed effect and \textit{Participant} as a random intercept. Positive values in the ``Difference'' column indicate higher perceived difficulty for Task 1 relative to Task 2. P values are Bonferroni corrected for multiple comparisons ($p_{\text{Bonferroni}}$).}
  \begin{tabular}{lclrrrrc}
    \toprule
    \multicolumn{3}{c}{Comparison} & \multicolumn{5}{c}{} \\
    \cmidrule(r){1-3}
    \textbf{Task 1} &  & \textbf{Task 2} & \textbf{Difference} & \multicolumn{1}{c}{\textbf{SE}} & \multicolumn{1}{c}{$\bm{t}$} & \multicolumn{1}{c}{$\bm{df}$} & $p_{\text{Bonferroni}}$ \\
    \midrule
    Act-Mot        & - & MA              & -2.439 & 0.183 & -13.304 & 567 & $< .001$ \\
    Act-Mot        & - & MI              & -0.606 & 0.182 &  -3.322 & 567 & 0.020   \\
    Act-Mot        & - & NB              & -4.172 & 0.183 & -22.824 & 567 & $< .001$ \\
    Act-Mot        & - & NB-MA           & -5.595 & 0.184 & -30.421 & 567 & $< .001$ \\
    Act-Mot        & - & NB-MA-Act-Mot   & -6.205 & 0.183 & -33.933 & 567 & $< .001$ \\
    Act-Mot        & - & PassMot         &  0.115 & 0.185 &   0.620 & 568 & 1.000   \\
    MA             & - & MI              &  1.834 & 0.183 &  10.024 & 567 & $< .001$ \\
    MA             & - & NB              & -1.733 & 0.183 &  -9.452 & 567 & $< .001$ \\
    MA             & - & NB-MA           & -3.156 & 0.184 & -17.108 & 567 & $< .001$ \\
    MA             & - & NB-MA-Act-Mot   & -3.765 & 0.183 & -20.532 & 567 & $< .001$ \\
    MA             & - & PassMot         &  2.554 & 0.186 &  13.749 & 568 & $< .001$ \\
    MI             & - & NB              & -3.567 & 0.182 & -19.560 & 567 & $< .001$ \\
    MI             & - & NB-MA           & -4.989 & 0.183 & -27.196 & 567 & $< .001$ \\
    MI             & - & PassMot         &  0.721 & 0.185 &   3.900 & 568 & 0.002   \\
    NB             & - & NB-MA           & -1.423 & 0.184 &  -7.736 & 567 & $< .001$ \\
    NB             & - & PassMot         &  4.287 & 0.185 &  23.145 & 568 & $< .001$ \\
    NB-MA          & - & PassMot         &  5.710 & 0.186 &  30.641 & 568 & $< .001$ \\
    NB-MA-Act-Mot  & - & MI              &  5.599 & 0.182 &  30.698 & 567 & $< .001$ \\
    NB-MA-Act-Mot  & - & NB              &  2.032 & 0.183 &  11.114 & 567 & $< .001$ \\
    NB-MA-Act-Mot  & - & NB-MA           &  0.609 & 0.184 &   3.313 & 567 & 0.021   \\
    NB-MA-Act-Mot  & - & PassMot         &  6.320 & 0.185 &  34.108 & 568 & $< .001$ \\
    \bottomrule
  \end{tabular}
  \label{tab:model_results_post-hoc}%
\end{table}

\begin{figure}[htbp]
\centering
\includegraphics[width=0.6\textwidth]{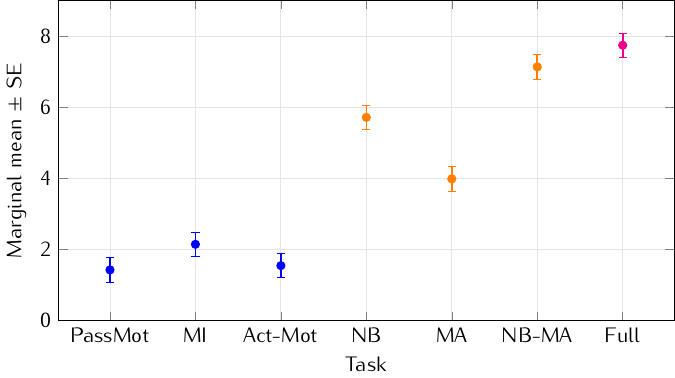}
\caption{Marginal mean and standard error for cognitive load rating of the 7 tasks.}
\label{fig:task_difficulty}%
\end{figure}

The model showed substantial explanatory power (conditional $R^2 = 0.81$, marginal $R^2 = 0.76$), with a significant main effect of Task on perceived difficulty ($F(6,567) = 418$, $p < 0.001$) and no significant effect of session ($F(2,568) = 0.324$, $p = 0.723$). Post hoc comparisons with Bonferroni correction indicated significant differences between all tasks ($p < 0.05$), except between Act-Mot and Pass-Mot (see Table~\ref{tab:model_results_post-hoc}). Overall, multitask conditions (NB-MA, Full) were rated as more demanding than single-component tasks. Figure~\ref{fig:task_difficulty} shows the statistics of task difficulty scores across participants.

\subsection{Neurophysiological validation - Representational Similarity Analysis}

\subsubsection{RSA approach}

To assess whether neural activity reflected differences in task difficulty, we used representational similarity analysis (RSA) \cite{kriegeskorte2008representational} on EEG and fNIRS data. RSA characterises each condition by its pattern of activity and summarises pairwise dissimilarities between conditions in a representational dissimilarity matrix (RDM). We then tested whether these neural RDMs were related to a theoretical RDM derived from participants' self-reported task difficulty.

For EEG and fNIRS, observations were defined as single-trial responses. Across the seven experimental conditions, we obtained 18 trials per condition, yielding 126 trials in total. For each participant and modality, we computed trial-by-trial similarity matrices using Pearson correlations between the 126 trial patterns. Correlation coefficients were Fisher z-transformed, averaged across participants and converted back to correlation values at the group level; dissimilarity was defined as $1 - r$ to obtain neural RDMs. This procedure was applied at the whole-head level (combining all EEG channels or all fNIRS channels) and at the channel level (separate RDMs for each EEG and fNIRS channel).

To relate neural representations to perceived task difficulty, we constructed a theoretical RDM from subjective difficulty ratings. For each participant, we computed a $7 \times 7$ dissimilarity matrix using Euclidean distances between mean difficulty ratings for each task and then averaged these matrices across participants to obtain a group-level difficulty model. Neural RDMs (EEG and fNIRS, whole-head and channel-wise) were compared to this theoretical RDM using Spearman correlations. Statistical significance was assessed using one-sided permutation tests with 5,000 permutations, in which condition labels were shuffled to generate a null distribution of correlation values. $p$-values (threshold $p < 0.05$) were corrected for multiple comparisons across channels using the false discovery rate (FDR) \cite{benjamini1995controlling}. 

\subsubsection{RSA results}

At the whole-head level, we did not observe a significant correspondence between neural and theoretical representations. The whole-head EEG RDM was not significantly correlated with the difficulty model (Spearman $\rho = -0.076$, $p = 0.9972$) and neither oxyhaemoglobin nor deoxyhaemoglobin whole-head fNIRS RDMs showed a significant correlation (O2Hb: $\rho = -0.011$, $p = 0.136$; HHb: $\rho = -0.011$, $p = 0.138$).

\begin{figure}[htb]
\centering
\includegraphics[width=\textwidth]{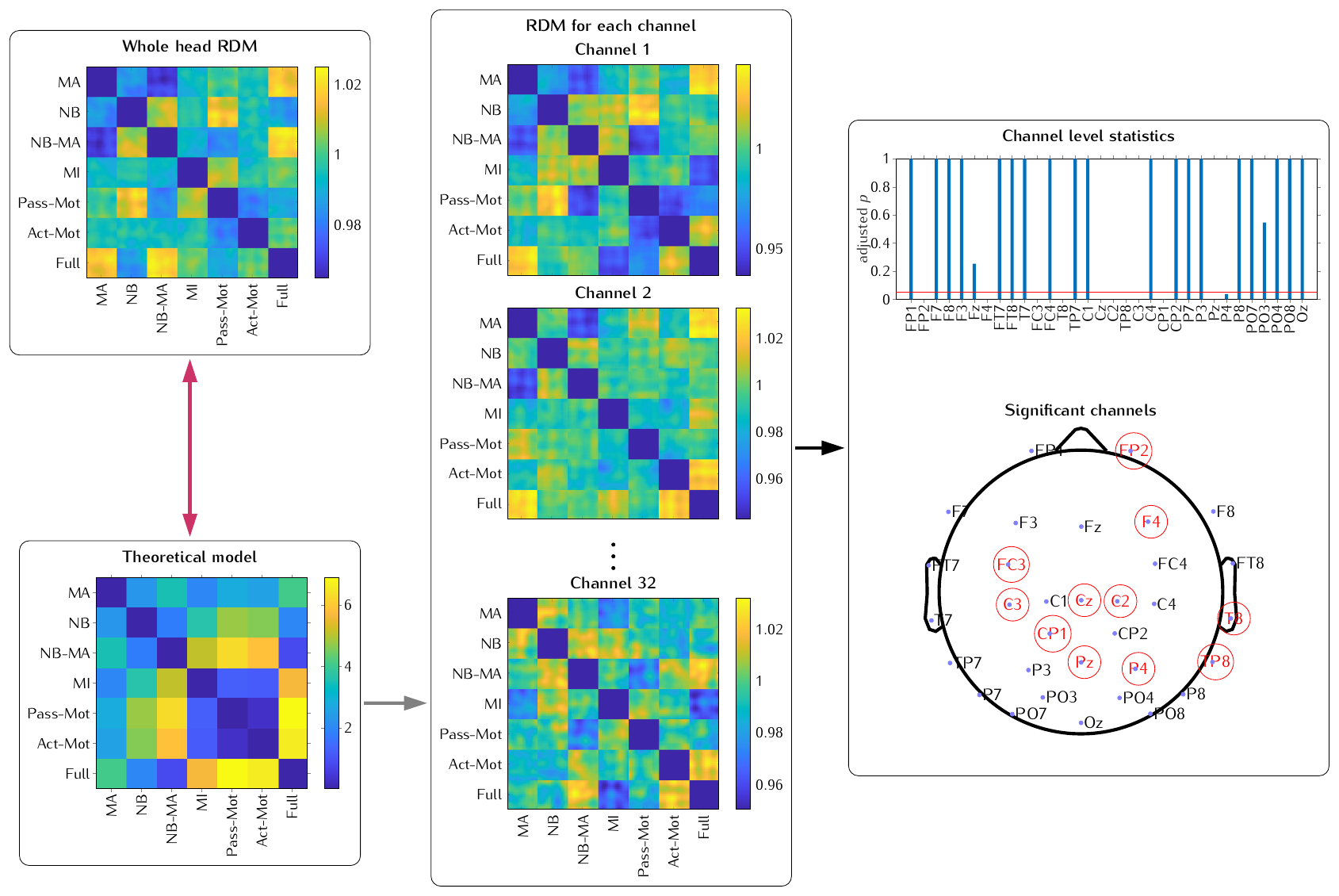}
\caption{EEG RSA pipeline and channel-level results. Left: group-level whole-head representational dissimilarity matrix (RDM) across the seven tasks. Centre: example channel-wise RDMs illustrating variability in representational structure across EEG channels. Right, top: channel-wise Spearman correlations between EEG RDMs and the task-difficulty model, with the FDR threshold indicated. Right, bottom: scalp topography of channels that shows significant correlations with the difficulty model after FDR correction.}
\label{fig:RSA_results_EEG}
\end{figure}

In contrast, channel-level analyses revealed condition-related structure in the EEG data. Several channels showed significant positive correlations between their neural RDMs and the difficulty model after FDR correction, primarily over frontal, central and temporal areas (i.e. FP2, Fz, F4, FC3, T8, Cz, C2, TP8, C3, CP1, Pz, P4; see Table~\ref{tab:EEG_channel_level_stats}). Figure~\ref{fig:RSA_results_EEG} illustrates the analysis pipeline and the topographic distribution of channel-wise RSA effects. These results indicate that, under the present preprocessing pipeline, EEG patterns at specific scalp sites carry information consistent with participants' perceived task difficulty, even though whole-head aggregation does not detect this structure.

\begin{table}[htbp]
\small
  \centering
  \caption{Spearman correlations ($\rho$) of EEG channels in RSA, channel with significant adjusted $p$-values are in bold.}
    \begin{tabular}{lc|lc|lc|lc}
    \toprule
    \textbf{Channel} & \textbf{$\rho$} & \textbf{Channel} & \textbf{$\rho$} & \textbf{Channel} & \textbf{$\rho$ } & \textbf{Channel} & \textbf{$\rho$ } \\ \midrule
    FP1   & 0.3401 & FT8   & 0.3234 & \hchan{C2} & 0.3854 & \hchan{Pz} & 0.4396 \\
    \hchan{FP2} & 0.5505 & T7    & 0.2713 & \hchan{TP8} & 0.4174 & \hchan{P4}    & 0.3716 \\
    F7    & 0.3157 & \hchan{FC3} & 0.4337 & \hchan{C3} & 0.4049 & P8    & 0.2341 \\
    F8    & 0.3337 & FC4   & 0.2646 & C4    & 0.2739 & PO7   & 0.2780 \\
    F3    & 0.2318 & \hchan{T8} & 0.4547 & \hchan{CP1} & 0.3836 & PO3   & 0.3598 \\
    \hchan{Fz} & 0.3646 & TP7   & 0.2983 & CP2   & 0.2076 & PO4   & 0.1614 \\
    \hchan{F4} & 0.4700 & C1    & 0.1538 & P7    & 0.2335 & PO8   & 0.1885 \\
    FT7   & 0.3031 & \hchan{Cz} & 0.3890 & P3    & 0.3230 & Oz    & 0.2209 \\ \bottomrule
    \end{tabular}
  \label{tab:EEG_channel_level_stats}%
\end{table}%

For fNIRS, channel-level analyses did not yield significant correlations with the difficulty model after FDR correction for either O2Hb or HHb signals (see supplementary Table~\ref{tab:O2Hb_HHb_stats}). A small number of channels showed weak, uncorrected correlations (for example, O2Hb channels 4 and 16), but these effects did not survive correction and should be interpreted cautiously. Overall, the present RSA results suggest that EEG channels provide clearer task-difficulty-related structure than fNIRS channels under the current analysis settings, while still indicating that fNIRS signals are of sufficient quality to support more targeted or alternative analyses in future work.

\section{Usage Notes}
\label{sec:usage_notes}

The dataset is publicly available on the OpenNeuro platform \cite{ds007554}. Data follow the BIDS specification \cite{gorgolewski2016brain}. For each participant (\ttb{sub-XXX}) and session (\ttb{ses-01}, \ttb{ses-02}, \ttb{ses-03}), modality-specific data are stored in \ttb{beh}, \ttb{eeg} and \ttb{nirs} folders, with corresponding JSON sidecars and event files. Recordings are provided before preprocessing, with associated metadata, so users can apply their own analysis pipelines.

\subsection{Software compatibility and basic usage}

EEG data are stored as \ttb{.edf} files with associated \ttb{\_channels.tsv} and \ttb{\_events.tsv} files. These can be imported directly into standard EEG toolboxes such as EEGLAB \cite{delorme2004eeglab}, FieldTrip \cite{Oostenveld2011FieldTrip}, MNE \cite{GramfortEtAl2013a}, or Brainstorm using corresponding BIDS import functionalities. 

fNIRS data are provided as \ttb{.snirf} files with associated \ttb{\_nirs.json}, \ttb{\_channels.tsv} and \ttb{\_events.tsv} files. These can be processed with packages such as Homer3 \cite{huppert2009homer}, the NIRS Toolbox (\url{https://github.com/huppertt/nirs-toolbox}), MNE-nirs (\url{https://github.com/mne-tools/mne-nirs}) or other software that supports SNIRF and BIDS-NIRS. 

Physiological and behavioural time series (ECG, push-button and Biodex traces) are stored in the \ttb{beh} folder as compressed tab-separated value files (\ttb{\_physio.tsv.gz}) with JSON sidecars. These can be read in any general data analysis environment. Subjective measures (Karolinska Sleepiness Scale and task difficulty ratings) are provided as tabular files that can be linked to sessions and tasks via subject and task labels.

\subsection{Event timing and multimodal alignment}

Task events are encoded in the \ttb{\_events.tsv} files associated with the EEG and fNIRS modalities (and, for response-related signals, can be inferred from push-button traces in the \ttb{beh} folder). These event files provide onsets, durations and condition labels for each task. Supplementary Table~\ref{tab:event_labels} summarizes the main \ttb{trial\_type} labels observed in the representative event files.

Within each modality, event onsets are defined relative to the start of the corresponding recording. For unimodal analyses, these event files can therefore be used directly to epoch EEG or fNIRS data. For multimodal analyses that combine EEG, fNIRS and physiological signals, users should ensure consistent use of task definitions and event codes across modalities.

The provided metadata and event files allow users to choose the filters, artefact handling, temporal realignment or resampling strategies required for their multimodal analyses.

\subsection{Practical considerations for reuse}

Several practical points are important for reuse:
\begin{itemize}
    \item \textbf{Participant coverage.} Participants are labelled \ttb{sub-001} to \ttb{sub-030}. Data from \ttb{sub-008} were excluded due to incomplete sessions. For \ttb{sub-010} to \ttb{sub-014}, EEG recordings were not usable because of technical problems. Their EEG data are therefore not provided, but fNIRS, physiological and behavioural modalities remain available. {Participants \ttb{sub-017} and \ttb{sub-019} were excluded from the EEG technical validation after quality-control checks, but this concerns only the validation pipeline and not the complete repository. Users should consult Supplementary Table~\ref{tab:data_completeness_summary} before defining analysis cohorts.}
    
    \item \textbf{Sampling differences in fNIRS.} fNIRS data for \ttb{sub-001} to \ttb{sub-020} were acquired at 10 Hz, whereas \ttb{sub-021} to \ttb{sub-030} were recorded at 50 Hz. The shared \ttb{.snirf} files retain these original sampling rates. In the technical validation analyses, 50 Hz data were resampled to 10 Hz for uniform processing; users may choose their own resampling strategy depending on the analysis.
    
    \item \textbf{Event timing.} Task events are encoded in the \ttb{\_events.tsv} files in both the \ttb{eeg} and \ttb{nirs} folders. These files provide onsets, durations and condition labels for each task and can be used to epoch EEG, fNIRS and physiological signals in a consistent way across modalities.
    
    \item \textbf{Preprocessing choices.} The EEG and fNIRS preprocessing pipelines described in the Technical Validation section (PREP and ASR for EEG; SCI-based pruning, wavelet motion correction, band-pass filtering and MBLL for fNIRS) were applied only for the validation analyses reported in this article. They are not imposed on users and should be considered as one possible starting point rather than a recommended standard pipeline. {For EEG analyses, users may also apply other ocular-artefact correction or mobile-EEG cleaning approaches if they are more adapted to their question \cite{Ronca2024o}.}

    \item {\textbf{MATLAB padding.} The released files keep the original signal lengths and sampling rates. For MATLAB workflows that require equal-length arrays, users may apply zero-padding or NaN-padding after loading the data. Such padding is an analysis-specific preprocessing step and is not part of the shared recordings.}
\end{itemize}

The combination of raw EEG, fNIRS, physiological and behavioural data across a hierarchy of cognitive, motor and combined tasks makes the dataset suitable for a broad range of applications. These include but are not limited to, testing alternative preprocessing pipelines, developing and benchmarking multimodal fusion methods, evaluating artefact removal and performing multivariate analyses such as decoding or RSA under controlled manipulations of task demands.

\subsection{Study limitations and scope of technical validation}

Several points should be considered when reusing this dataset. First, the cohort includes healthy adults only and should not be interpreted as a clinical or population-normative sample. Second, the sample size is moderate, and the number of usable recordings differs across modalities because some recordings are missing or excluded from specific validation analyses. Third, the tasks were performed in a controlled laboratory setting. They are more complex than isolated single-task paradigms, but they do not reproduce fully natural daily-life conditions. Fourth, standardized workload and perceived-exertion scales, such as NASA-TLX or Borg RPE \cite{Hart2006Nasa,Borg1990Psychophysical}, were not collected. Finally, ECG is provided as a synchronized auxiliary stream for future reuse but was not included in the present technical validation; ECG-specific analyses therefore require dedicated quality control and preprocessing.

\section*{Author Contribution}

Study design: G.V. Data acquisition: Z.A., G.V., L.E. and S.Pl. Data analysis: Z.A. and G.V. Drafting of the manuscript: Z.A., G.V. and B.X. Supervision: S.P., G.D. and J.M. All authors reviewed and revised the manuscript and approved the final version.

\section*{Funding}

Z.A. received a doctoral fellowship from AXIAUM University of Montpellier-ISDM (ANR-20-THIA-0005-01) and Doctoral School in ICT, Mathematics, Physics and Engineering (ED I2S) in France.\\
G.V. was supported by the Occitanie Region, France, through the SKYPHYSIA project.

\section*{Competing Interests}

The authors have no conflicting interests.

\section{Code Availability}

The shared dataset consists of raw recordings only, so users are free to apply their preferred preprocessing and analysis pipelines. The technical validation analyses reported in this article were implemented in MATLAB R2021b using publicly available toolboxes (EEGLAB \cite{delorme2004eeglab}, PREP \cite{bigdely2015prep}, clean\_rawdata, Homer3 \cite{huppert2009homer} and qt-nirs), together with custom MATLAB scripts for data formatting, preprocessing and representational similarity analysis.

\section{Data Availability}

This dataset is organised according to the BIDS standard \cite{gorgolewski2016brain} and was validated using the online BIDS validator. It is available from the open-access repository OpenNeuro under DOI \url{https://doi.org/10.18112/openneuro.ds007554.v1.0.0} \cite{ds007554}. Participants are identified as \ttb{sub-001} to \ttb{sub-030}. For each available session (\ttb{ses-01}, \ttb{ses-02}, \ttb{ses-03}), we provide raw EEG, fNIRS, physiological, behavioural and subjective data for the seven experimental tasks. No preprocessed (derivative) data are included, so that users can apply their own analysis pipelines.

\bibliographystyle{IEEEtranDOI}
\bibliography{bib}

\clearpage
\linespread{1}
\appendix
\makeatletter
\renewcommand \thesection{S\@arabic\c@section}
\renewcommand\thetable{S\@arabic\c@table}
\renewcommand \thefigure{S\@arabic\c@figure}
\makeatother
\setcounter{section}{0}
\setcounter{table}{0}
\setcounter{figure}{0}

\begin{table}[!ht]
\footnotesize
\begin{center}
\caption{Summary of known participant and modality availability. This table separates repository availability from inclusion in the technical validation analyses.}
\label{tab:data_completeness_summary}
\begin{tabular}{p{4cm}p{3.5cm}p{7cm}}
\toprule
\textbf{Participant(s)} & \textbf{Availability status} & \textbf{Implication for reuse} \\
\midrule
\ttb{sub-001}--\ttb{sub-007}, \ttb{sub-009}, \ttb{sub-015}--\ttb{sub-016}, \ttb{sub-018}, \ttb{sub-020}--\ttb{sub-030} & Released modalities available according to repository structure & Eligible for modality-specific analyses after user-defined quality control. \\ \midrule
\ttb{sub-008} & Excluded from released dataset & Recording sessions were incomplete. \\ \midrule
\ttb{sub-010}--\ttb{sub-014} & EEG unavailable; fNIRS, physiological and behavioural data available & Exclude from analyses requiring EEG; may be used for fNIRS, ECG, behavioural, or subjective analyses. \\ \midrule
\ttb{sub-017}, \ttb{sub-019} & Retained in repository; excluded from EEG technical validation & Distinguish repository availability from inclusion in the validation pipeline. \\
\bottomrule
\end{tabular}
\end{center}
\end{table}

\begin{table}[!ht]
\small
\centering
\caption{Spearman correlations, raw $p$-values and adjusted $p$-values for fNIRS channels}
\label{tab:O2Hb_HHb_stats}
\begin{minipage}[t]{0.42\textwidth}
\centering
\begin{tabular}{c|c|c|c}
\toprule
\multicolumn{4}{c}{\textbf{O2Hb}} \\ \midrule
\textbf{Channel} & \textbf{$\rho$ Spearman} & \textbf{adj\_$p$} & \textbf{raw\_$p$} \\ \midrule
1  & 0.0126 & 0.4442 & 0.1388 \\ \hline
2  & 0.0006 & 0.5543 & 0.4850 \\ \hline
3  & 0.0159 & 0.3248 & 0.0812 \\ \hline
4  & 0.0195 & 0.2795 & 0.0396 \\ \hline
5  & 0.0022 & 0.5228 & 0.4140 \\ \hline
6  & -0.0020 & 0.5800 & 0.5800 \\ \hline
7  & 0.0037 & 0.5228 & 0.3722 \\ \hline
8  & 0.0035 & 0.5228 & 0.3784 \\ \hline
9  & 0.0183 & 0.2795 & 0.0524 \\ \hline
10 & 0.0085 & 0.4500 & 0.2228 \\ \hline
11 & 0.0082 & 0.4500 & 0.2250 \\ \hline
12 & -0.0024 & 0.5800 & 0.5730 \\ \hline
13 & 0.0021 & 0.5228 & 0.4248 \\ \hline
14 & 0.0071 & 0.4619 & 0.2598 \\ \hline
15 & 0.0101 & 0.4500 & 0.1840 \\ \hline
16 & 0.0209 & 0.2795 & 0.0308 \\ \bottomrule
\end{tabular}
\end{minipage}
\hfill
\begin{minipage}[t]{0.42\textwidth}
\centering
\begin{tabular}{c|c|c|c}
\toprule
\multicolumn{4}{c}{\textbf{HHb}} \\ \midrule
\textbf{Channel} & \textbf{$\rho$ Spearman} & \textbf{adj\_$p$} & \textbf{raw\_$p$} \\ \midrule
1  & 0.0008  & 0.9723 & 0.4696 \\ \hline
2  & -0.0070 & 0.9723 & 0.7292 \\ \hline
3  & 0.0047  & 0.9723 & 0.3442 \\ \hline
4  & -0.0135 & 0.9888 & 0.8840 \\ \hline
5  & 0.0090  & 0.9723 & 0.2194 \\ \hline
6  & -0.0259 & 0.9888 & 0.9888 \\ \hline
7  & -0.0214 & 0.9888 & 0.9736 \\ \hline
8  & -0.0229 & 0.9888 & 0.9776 \\ \hline
9  & 0.0091  & 0.9723 & 0.2036 \\ \hline
10 & 0.0050  & 0.9723 & 0.3454 \\ \hline
11 & -0.0028 & 0.9723 & 0.6064 \\ \hline
12 & -0.0046 & 0.9723 & 0.6584 \\ \hline
13 & 0.0026  & 0.9723 & 0.4108 \\ \hline
14 & -0.0053 & 0.9723 & 0.6932 \\ \hline
15 & -0.0059 & 0.9723 & 0.6928 \\ \hline
16 & 0.0183  & 0.8768 & 0.0548 \\ \bottomrule
\end{tabular}
\end{minipage}
\end{table}

\begin{table}[ht]
\centering
\caption{Event-label logic for the \texttt{events.tsv} files. These files contain the columns \texttt{onset}, \texttt{duration}, and \texttt{trial\_type}. Onsets and durations are expressed in seconds. In representative files, the observed \texttt{trial\_type} labels are \texttt{start}, \texttt{target}, and \texttt{trigger}.}
\label{tab:event_labels}
\small
\begin{tabularx}{\linewidth}{l l X}
\toprule
\textbf{Task file} & \textbf{Observed \texttt{trial\_type} label} & \textbf{Interpretation} \\
\midrule
\texttt{activemotor} & \texttt{start}, \texttt{target} & \texttt{start} indicates the beginning of the task sequence. \texttt{target} indicates active motor cues or trials. \\ \midrule

\texttt{passivemotor} & \texttt{start}, \texttt{target} & \texttt{start} indicates the beginning of the task sequence. \texttt{target} indicates passive motor cues or trials. \\ \midrule

\texttt{motorimagery} & \texttt{target} & \texttt{target} indicates motor imagery cues or trials. No \texttt{start} event was present in the inspected representative file. \\ \midrule

\texttt{nback} & \texttt{start}, \texttt{trigger}, \texttt{target} & \texttt{start} indicates the beginning of the task sequence. \texttt{target} indicates target N-back stimuli. \texttt{trigger} indicates non-target N-back stimuli. \\ \midrule

\texttt{mentalarithmetic} & \texttt{start}, \texttt{target} & \texttt{start} indicates the beginning of the task sequence. \texttt{target} indicates mental arithmetic stimuli or trials. \\ \midrule

\texttt{nbackarithmetic} & \texttt{start}, \texttt{trigger}, \texttt{target} & \texttt{start} indicates the beginning of the task sequence. \texttt{target} indicates target trials. \texttt{trigger} indicates non-target or intermediate stimulus events. \\ \midrule

\texttt{full} & \texttt{start}, \texttt{trigger}, \texttt{target} & \texttt{start} indicates the beginning of the task sequence. \texttt{target} and \texttt{trigger} follow the same target/non-target logic as the combined cognitive task, here embedded in the full cognitive-motor condition. \\
\bottomrule
\end{tabularx}
\end{table}

\end{document}